\documentclass[10pt,a4paper,twocolumn]{article}
\usepackage[a4paper,left=16mm,right=16mm,bottom=25mm,top=25mm]{geometry}
\usepackage[switch]{lineno} 
\usepackage{lipsum} 
\usepackage{footmisc}  
\usepackage[fleqn]{amsmath}
\usepackage{fancyhdr}
\usepackage{subcaption}
\usepackage{graphicx}
\usepackage{booktabs}
\usepackage[british]{babel}
\usepackage[square,comma,numbers,sort&compress]{natbib}
\usepackage{comment}
\usepackage{lettrine}

\usepackage{csvsimple}
\usepackage{graphicx}
\usepackage{subcaption}
\usepackage{amsmath}
\usepackage{tikz}
\usepackage{array}
\usepackage{tabu}
\usepackage{multirow}
\usepackage{url}
\usepackage{xcolor}
\usepackage{afterpage}
\usepackage{graphicx}%
\usepackage{bm}%
\usepackage[title]{appendix}
\usepackage{float}
\usepackage{hyperref}
\hypersetup{colorlinks,linkcolor={blue},citecolor={blue},urlcolor={cyan}}
\usepackage{cleveref}
\usepackage{afterpage}
\usepackage{caption}
\usepackage{lineno}
\usepackage{lettrine}
\usepackage{color}
\usepackage{soul}
\usepackage{ulem}
\usepackage{float}
\usepackage{placeins}
\usepackage{wrapfig}

\newcommand{\bluemark}[1] {\color{blue}{#1}\color{black}\normalsize}
\usepackage{subcaption}

\definecolor{greendark}{rgb}{0.0,0.5,0.0}
\definecolor{darkred}{rgb}{0.55, 0.0, 0.0}
\definecolor{navy}{rgb}{0.0, 0.0, 0.5}
\usepackage{graphicx}
\usepackage{subcaption}

\title{Chiral Magnons and Cycloidal Phonons in Altermagnetic CuF$_{2}$ Monolayer}

\author{Andrea M. León$^{1}$, Mat\'{\i}as F. Torreblanca$^{1}$, Carmine Autieri$^{2}$, Jhon W. González$^{3}$}
\date{
$^{1}$Departamento de Física, Facultad de Ciencias, Universidad de Chile, Casilla 653, Santiago, Chile.\\
$^{2}$International Research Centre Magtop, Institute of Physics, Polish Academy of Sciences, Aleja Lotników 32/46, 02668 Warsaw, Poland.\\
$^{3}$Departamento de Física, Universidad de Antofagasta, Av. Angamos 601, Casilla 170, Antofagasta, Chile.\\[2ex]
}

\begin{document}
\twocolumn[
\begin{@twocolumnfalse}
\maketitle
\vspace{10pt} 
\noindent
\begin{small} 
\setlength{\parskip}{0pt} 
\setlength{\parindent}{0pt} 
\noindent
Altermagnetism establishes momentum-dependent spin splitting through 
non-symmorphic crystal symmetries, yet whether these same symmetries 
simultaneously govern spin and lattice collective excitations remains 
open. Here we show, using first-principles calculations and linear 
spin-wave theory, that monolayer CuF$_2$ hosts both chirality-split 
magnons and cycloidal phonons controlled by the same $P2_1/c$ 
symmetry operations. The altermagnetic order drives strongly 
anisotropic magnon chirality via symmetric anisotropic exchange, 
with Dzyaloshinskii--Moriya interactions acting as a weak secondary 
modulation. Crucially, the phonon and magnon chiral responses are 
directionally complementary: cycloidal phonon angular momentum 
emerges precisely where magnon chirality is symmetry-suppressed, 
and vice versa. The magnon bands further carry quantized Chern 
numbers $C^M = \pm 2$, confirming non-trivial altermagnetic 
topology. These results establish monolayer CuF$_2$ as a platform 
where a single symmetry framework engineers magnonic, phononic, 
and topological responses, providing a direct connection between altermagnetism and spin-lattice chirality in two-dimensional materials.

\end{small}
\vspace{15pt}

\end{@twocolumnfalse}]


\section*{Introduction}
\lettrine[lines=2, loversize=0.25, findent=2pt]{\LettrineFont{\bluemark{A}}}
ltermagnetism has recently been established as a symmetry-protected 
magnetic phase distinct from both ferromagnetism and antiferromagnetism, 
characterized by momentum-dependent spin splitting that arises without 
net magnetization and without relativistic spin-orbit coupling 
\cite{Smejkal22emerging,Smejkal22beyond,cheong2024altermagnetism}. 
Unlike conventional antiferromagnets, altermagnets host non-relativistic 
spin-split bands governed by rotational crystal symmetries, giving rise 
to phenomena previously associated only with ferromagnets, including 
anomalous Hall effects and orbital magnetization 
\cite{g32j-hnvz,PhysRevLett.134.196703,vzmh-mxlz,doi:10.1021/acs.jpclett.5c03677,tenzin2025persistent}. 
A natural and largely open question is whether these same symmetry principles extend beyond the electronic sector: can they also govern 
collective spin and lattice excitations, and do magnonic and phononic chirality share a common symmetry origin?

In the magnonic sector, the altermagnetic symmetry generates 
chirality-split magnon spectra, sometimes termed altermagnons 
\cite{issing2026altermagnons}, whose directional character directly 
reflects the underlying spin-space group, enabling non-relativistic 
control of magnonic responses \cite{vsmejkal2023chiral,gomonay2024structure}. 
Recent phenomenological work has shown that the key ingredient is a 
sublattice-dependent anisotropic spin stiffness, with the Dzyaloshinskii--Moriya interactions playing only a secondary role 
\cite{gomonay2024structure}. In the phononic sector, broken crystal 
or magnetic symmetries can endow lattice vibrations with finite angular 
momentum, producing chiral phonons \cite{juraschek2025chiral,zhang2026comprehensive} 
that couple to magnetic fields and spin degrees of freedom 
\cite{zhang2014angular,fransson2023chiral}, and manifest in the phonon 
Hall effect and anomalous thermal transport 
\cite{che2025magnetic,bao2026magnetic}. In conventional magnetic 
systems, however, such phonon chirality is typically weak and relies 
on large spin-orbit coupling \cite{thingstad2019chiral,yao2025theory}. 
Whether altermagnets, which generate spin splitting without spin-orbit coupling, can also impose finite phonon angular momentum 
through their non-symmorphic symmetry operations, and how magnonic and phononic chirality relate to each other in momentum space, remains largely unexplored; recent work has begun to address the magnonic sector through spin-point-group classifications 
\cite{zhang2026oddparity,wang2025alteraxial}, but the coupled 
spin--lattice response and its directional selectivity in 
collinear altermagnets remain open 
\cite{wang2023magnon,metzger2024magnon}.

\begin{figure}[t]
\centering
\includegraphics[width=0.5\textwidth]{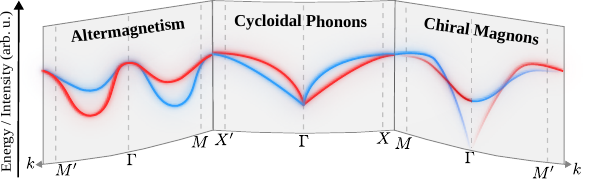}
\caption{Schematic representation of altermagnetic spin-splitting, cycloidal phonons and chiral magnons in CuF$_2$.}
\label{fig:0}
\end{figure}

Here, we address these questions by investigating monolayer CuF$_2$ 
\cite{PhysRevMaterials.8.034407}, recently identified as a 
$d$-wave altermagnet \cite{bandyopadhyay2026d,peng2025ferroelastic,peng2026sliding}. 
We demonstrate that the altermagnetic order generates strongly 
anisotropic chiral magnon dispersions, dominated by symmetric 
anisotropic exchange rather than by relativistic interactions, 
and that the system simultaneously hosts cycloidal phonon modes 
with finite angular momentum. Crucially, both responses are governed 
by the same non-symmorphic symmetry operations of the $P2_1/c$ 
spin space group, yet are directionally complementary: phonon chirality emerges precisely along the momentum directions where magnon chirality is suppressed by symmetry, and vice versa. 
This three-fold interplay between altermagnetic spin splitting, cycloidal phonons, and chiral magnons is summarized schematically 
in Fig.~\ref{fig:0}, and establishes monolayer CuF$_2$ as a platform 
where a single symmetry framework simultaneously leads to magnonic, phononic, and topological responses, providing a general principle 
for designing coupled spin--lattice chirality in two-dimensional 
quantum materials.

\footnotetext[1]{\textsuperscript{*} Corresponding author: andrea.leon@uchile.cl}
\footnotetext[2]{\textsuperscript{\dag} jhon.gonzalez@uantof.cl}

\section*{Methodology}

\subsection*{Computational Details}
We perform first-principles calculations with the Vienna Ab-initio Simulation Package (VASP)~\cite{VASP} using the projector augmented-wave  method~\cite{kresse1999ultrasoft}. Exchange-correlation effects are described within the generalized gradient approximation in the Perdew-Burke-Ernzerhof  parametrization~\cite{PBE}. We use a kinetic-energy cutoff of 500~eV for the plane-wave basis set, above the recommended values of the employed PAW pseudopotentials.  
To extract the magnetic exchange coupling constants, additional calculations are performed within a localized basis-set DFT framework using the OpenMX package~\cite{ozaki2004numerical,ozaki2005efficient} within equivalent parameters.

We model the CuF$_2$ monolayer in a slab geometry with a vacuum region of at least 15~\AA\ along the out-of-plane direction to prevent spurious interlayer interactions. The Brillouin zone is sampled with Monkhorst-Pack meshes of $10\times9\times1$ for structural relaxations and $20\times18\times1$ for static self-consistent calculations, corresponding to $k$-point spacings of approximately $0.02$ and $0.01$ $2\pi/$\AA{}, respectively~\cite{Monkorst}. We use Methfessel-Paxton smearing with a width of 0.05~eV during ionic relaxations. Atomic positions are relaxed until the residual forces on each atom are smaller than $10^{-3}$~eV/\AA, with electronic self-consistency reached within $10^{-6}$~eV.  
To account for strong on-site Coulomb interactions of Cu $4d$ states, we apply the rotationally invariant DFT+$U$ approach in the Dudarev formalism~\cite{dudarev1998electron}, with an effective parameter $U_{\mathrm{eff}} = 4$~eV. This choice, consistent with earlier studies of Cu fluorides~\cite{ZHENG20121703}, reproduces key experimental observables such as the insulating band gap and the magnitude of local magnetic moments in CuF$_2$.

\subsection*{Theoretical formalism}

\textit{Magnons}: We model the low-energy spin excitations in CuF$_2$ using a classical spin Hamiltonian, with parameters extracted from first-principles DFT+$U$ calculations performed with the \texttt{OpenMX} package~\cite{ozaki2004numerical,ozaki2005efficient}. The \texttt{TB2J} package~\cite{he2021tb2j} is used to obtain the isotropic Heisenberg couplings $J^{\rm iso}_{ij}$, the Dzyaloshinskii-Moriya interaction (DMI) vectors $\mathbf{D}_{ij}$, and the symmetric anisotropic exchange tensors $\mathbf J_{ij}^{\rm ani}$ for the Hamiltonian:
\begin{equation}
\mathcal{H} = \sum_{i<j}J^{\rm iso}_{ij}\,\mathbf S_i\cdot\mathbf S_j
+ \sum_{i<j}\mathbf D_{ij}\cdot(\mathbf S_i\times\mathbf S_j)
+ \sum_{i<j}\mathbf S_i\cdot\mathbf J_{ij}^{\rm ani}\cdot\mathbf S_j.
\end{equation}
Linear spin-wave theory (LSWT) is then applied using the \texttt{Magnopy} package~\cite{MAGNOPY,ivanov2021fast}, which constructs and diagonalizes the bosonic Bogoliubov-de Gennes (BdG) Hamiltonian, $\mathcal{H}_{\rm BdG}(\mathbf{k})$, yielding the magnon dispersion $\omega_n(\mathbf{k})$ and eigenvectors $\psi_n(\mathbf{k})$.

\begin{figure*}[ht] 
\subcaptionbox{}[0.34\textwidth]{%
\includegraphics[width=\linewidth]{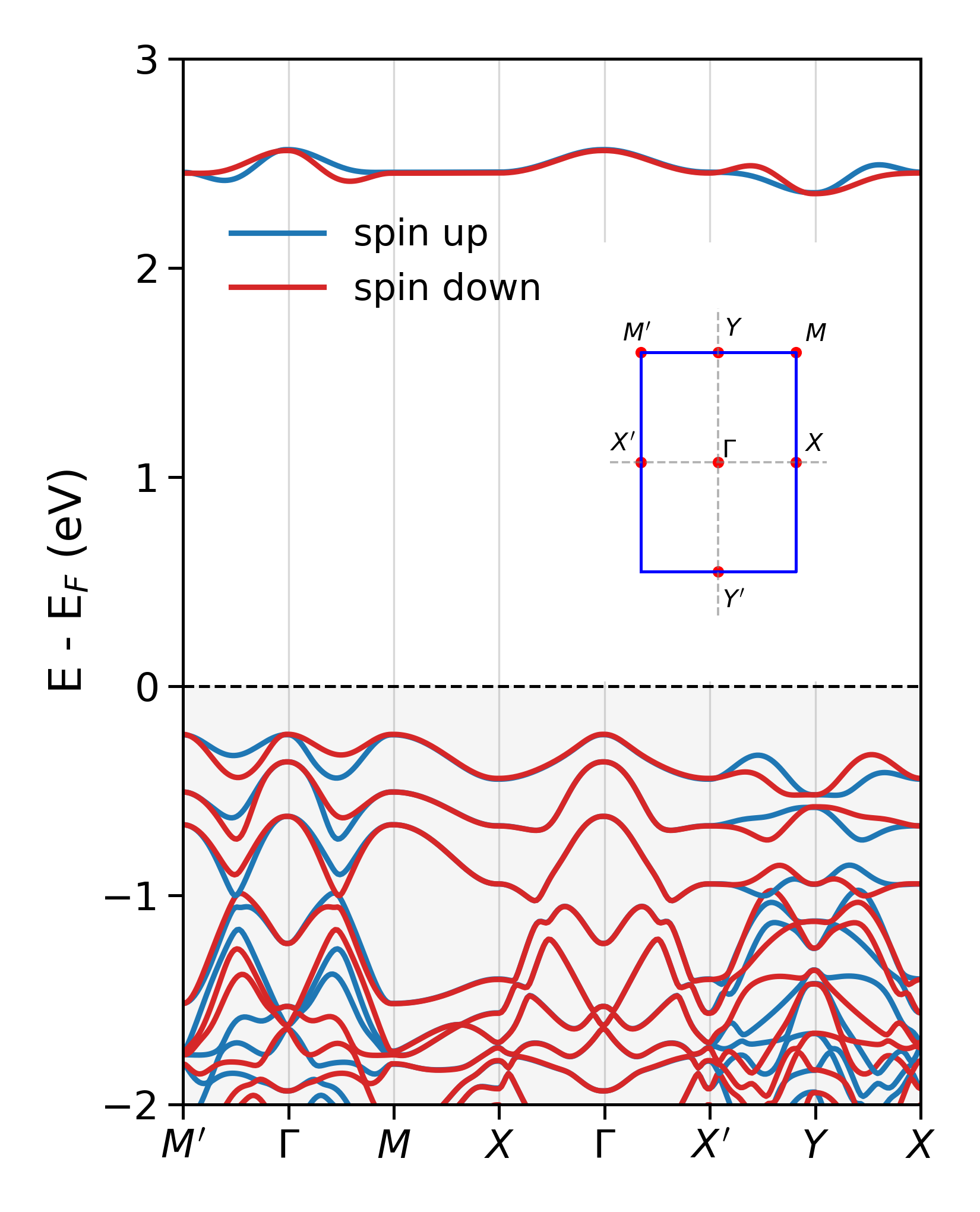}}
\subcaptionbox{}[0.33\textwidth]{%
\includegraphics[width=\linewidth]{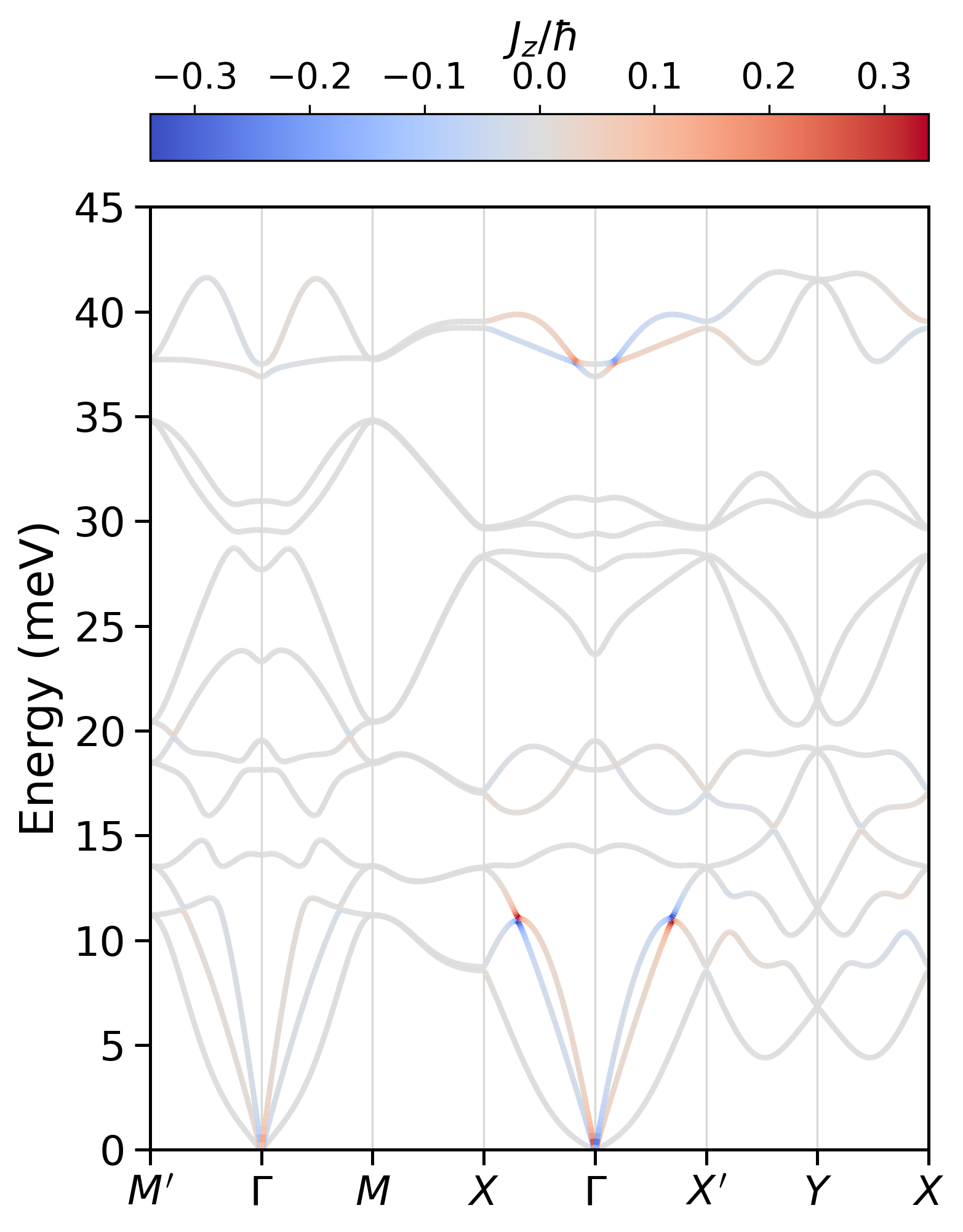}}
\subcaptionbox{}[0.33\textwidth]{%
\includegraphics[width=\linewidth]{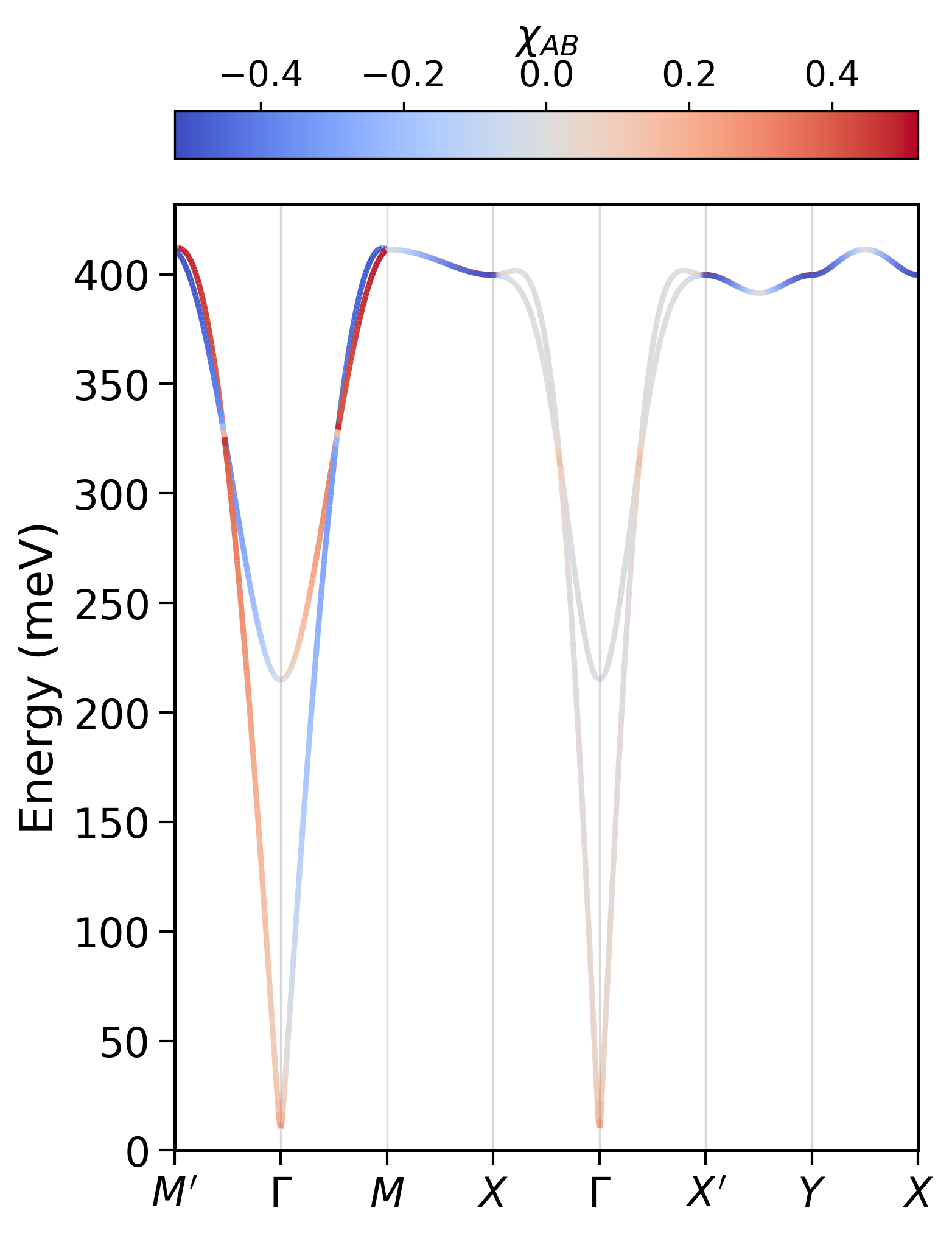}}
    \caption{(a) Electronic band structure highlighting the path with non-relativistic spin splittings and spin degeneracy. In the inset, we report the high-symmetry k-points of the two-dimensional Brillouin zone. 
(b) Phonon dispersion along with the mode-resolved phonon angular momentum. 
(c) Magnon band structure. The color scale represents the magnitude and sign of the phonon angular momentum $J_z$ and the inter-sublattice coherence $\chi_{AB}$, which quantifies the degree of hybridization between the $A$ and $B$ sublattices. Red (blue) indicates negative (positive) values of $\chi_{AB}$.}
    \label{fig:3}
\end{figure*}

To quantify the internal phase structure of these modes, we introduce the complex inter-sublattice correlator
\begin{equation}
\Xi_{AB}(\mathbf{k}) =
u_A^\dagger(\mathbf{k})\,u_B(\mathbf{k})
-
v_A^\dagger(\mathbf{k})\,v_B(\mathbf{k}),
\end{equation}
constructed from the particle and hole Bogoliubov amplitudes $\{u,v\}$ after projecting onto sublattices $A$ and $B$. The chiral inter-sublattice response plotted in Fig.~\ref{fig:3}(c) is defined as $\chi_{AB}(\mathbf{k}) \equiv \mathrm{Im}[\Xi_{AB}(\mathbf{k})],$ which captures the handed component associated with the relative phase between sublattices.

The topological character is quantified by the magnon Chern number,
\begin{equation}
C_n^M = \frac{1}{2\pi} \int_{\mathrm{BZ}} \Omega_{xy}^{(n)}(\mathbf{k}) \, d^2k,
\end{equation}
where the Berry curvature for band $n$ is defined as
\begin{equation}
\Omega_{xy}^{(n)}(\mathbf{k}) = i \left[ \langle \partial_{k_x} \psi_n | \tau_z | \partial_{k_y} \psi_n \rangle - \langle \partial_{k_y} \psi_n | \tau_z | \partial_{k_x} \psi_n \rangle \right],
\end{equation}
computed from the bosonic BdG eigenvectors $\psi_n(\mathbf{k})$ with paraunitary normalization $\langle \psi_n|\tau_z|\psi_n\rangle=1$~\cite{shindou2013topological}. Numerically, we evaluate the integral on a discrete $\mathbf{k}$-mesh using the gauge-invariant Fukui formulation~\cite{fukui2005chern}, adapted to the bosonic metric via link variables $U_\mu^{(n)}(\mathbf{k}) = \langle \psi_n(\mathbf{k})|\tau_z|\psi_n(\mathbf{k}+\hat{k}_\mu)\rangle / |\langle \psi_n(\mathbf{k})|\tau_z|\psi_n(\mathbf{k}+\hat{k}_\mu)\rangle|$. This procedure yields quantized Chern numbers for the magnon bands~\cite{AgF2_paper}.

\begin{figure}[ht] 
\centering
{\includegraphics[width=0.5\textwidth]{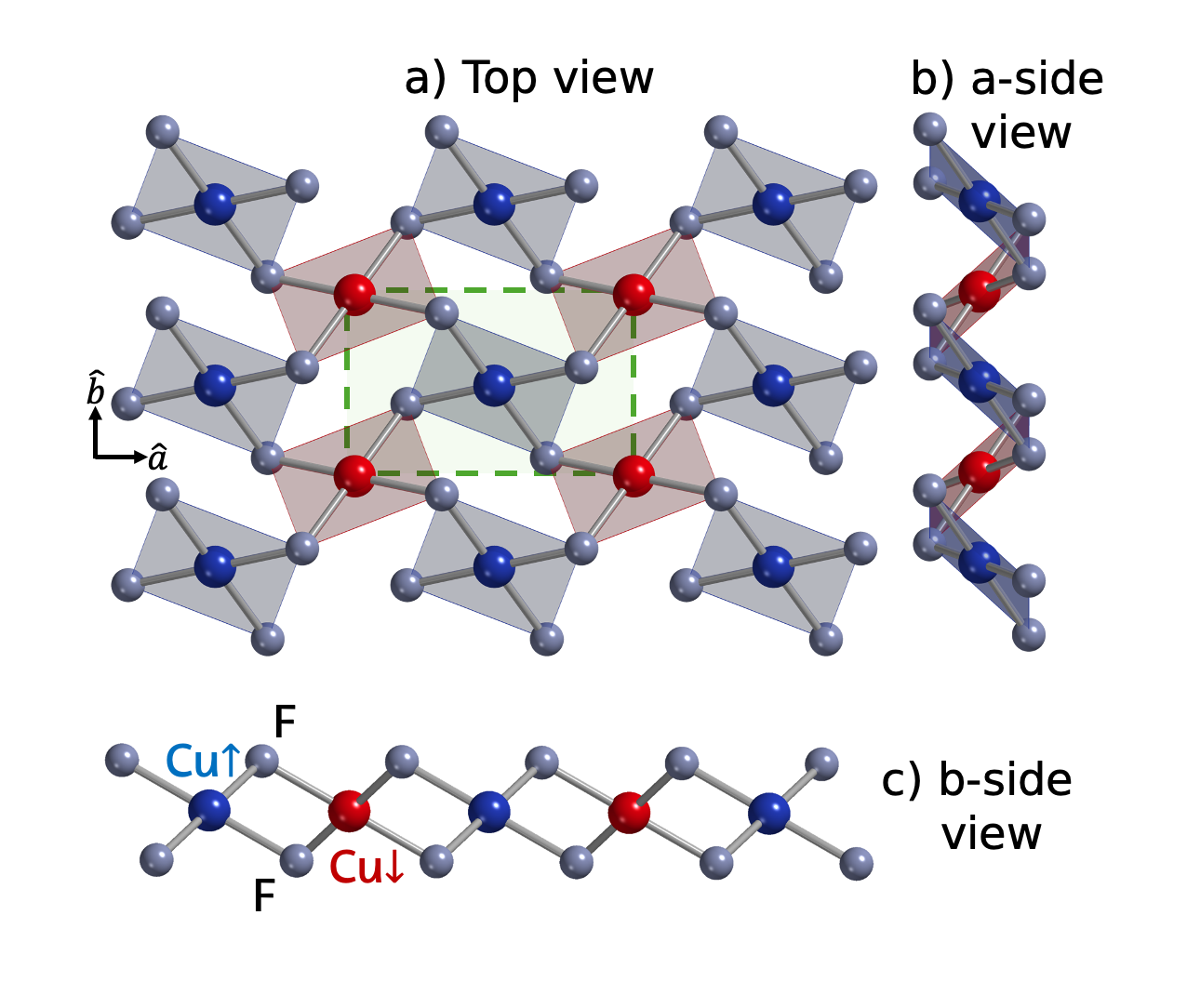}}
    \caption{(a) Top view of the CuF$_{2}$ monolayer. Panels (b) and (c) display side views along the $a$-$b$ plane. The dashed green line encloses the unit cell.}
    \label{fig:1}
\end{figure}

\textit{Phonons:} The harmonic force constants are computed using density functional perturbation theory (DFPT) and post-processed using the \textsc{Phonopy} package~\cite{togo2023implementation,togo2023first,phonopy-web} to obtain phonon frequencies and eigenvectors in a $5\times5\times1$ supercell. 
Phonon angular momentum was evaluated using an extended implementation of the \texttt{phonon\_angular\_momentum} script \cite{QijingZheng,zhang2014angular,zhang2015chiral}. Following Zhang and Niu~\cite{zhang2014angular}, the microscopic phonon angular momentum is defined as
\begin{equation}
  \mathbf{J}^{\mathrm{ph}} = \sum_{l,\kappa} \mathbf{u}_{l\kappa} \times \dot{\mathbf{u}}_{l\kappa},
\end{equation}
where the summation runs over all unit cells $l$ and atoms $\kappa$, and $\mathbf{u}_{l\kappa}$ are the mass-weighted atomic displacements. 
For a harmonic normal mode $\nu$ at wavevector $\mathbf{q}$, the atomic displacements can be written as $\mathbf{u}_{l\kappa}(t) \propto \mathbf{u}_{\kappa\mathbf{q}\nu} e^{-i\omega_{\mathbf{q}\nu}t}$. Substituting this form into the microscopic definition and performing a time average yields the mode-resolved phonon angular momentum
\begin{equation}
  \mathbf{J}_{\mathbf{q}\nu} = \hbar \sum_{\kappa} \mathrm{Im}\left( \mathbf{u}^{*}_{\kappa\mathbf{q}\nu} \times \mathbf{u}_{\kappa\mathbf{q}\nu} \right),
  \label{Eq:J}
\end{equation}
where $\mathbf{u}_{\kappa\mathbf{q}\nu}$ are the mass-weighted complex polarization vectors normalized such that $\sum_\kappa |\mathbf{u}_{\kappa\mathbf{q}\nu}|^2 = 1$ \cite{zhang2014angular}.
All phonon calculations are performed without spin-orbit coupling.

\section*{Results}



Bulk CuF$_2$ is a correlated insulating antiferromagnet, commonly described as a Mott-Hubbard-type insulator with a gap associated with localized Cu $3d$ states \cite{ZHENG20121703}. Optical measurements report low-energy transitions in the range of 1.2--1.93~eV \cite{aramburu2019explaining,oelkrug2008absorption}, while first-principles estimates of the fundamental gap depend sensitively on the treatment of electronic correlations, in particular on the chosen Hubbard $U$ parameter. Crucially, while the gap magnitude varies with $U$, the $d$-wave altermagnetic spin splitting remains qualitatively robust across the physically relevant range $U_{\rm eff} = 3$--6~eV~\cite{bandyopadhyay2026d}. This robustness underpins our choice of $U_{\rm eff} = 4$~eV for the monolayer calculations, which simultaneously reproduces the insulating character and the local magnetic moment discussed below.

The Cu$^{2+}$ ions exhibit a $3d^9$ electronic configuration, with one unpaired hole that would nominally correspond to an atomic magnetic moment of $1~\mu_B$. In our calculation, we obtain a magnetic moment of $\approx 0.75~\mu_B$ per Cu atom within the chosen Wigner-Seitz radius of $1.16~$\AA. Structurally, bulk CuF$_2$ crystallizes in the monoclinic space group $P2_1/c$ (No.~14), where Cu atoms occupy symmetry-equivalent Wyckoff positions and are coordinated by four fluorine atoms forming distorted CuF$_4$ plaquettes. Consequently, in the nonmagnetic crystallographic structure, the two Cu sublattices are related by an inversion center and are therefore crystallographically equivalent.
%
%
The situation changes once the collinear AFM Néel order is imposed. The inversion operation maps one Cu sublattice onto the other, but the two sites carry opposite magnetic moments. Since spatial inversion does not reverse the spin direction, the magnetic configuration is not invariant under inversion alone. Thus, the Néel order breaks the pure inversion symmetry of the magnetic structure, even though the underlying ionic lattice remains centrosymmetric. Equivalently, the inversion-related Cu sites become magnetically inequivalent because they belong to opposite spin sublattices, as shown in Figs.~\ref{fig:1}(a)--(c). This selective breaking of inversion is the prerequisite that allows the system to retain only combined spin-space symmetries, which in turn govern the momentum-dependent altermagnetic spin splitting and the complementary chiral responses discussed below.

In the two-dimensional limit, earlier studies have proposed that the monolayer preserves the bulk crystal symmetry \cite{peng2025ferroelastic}. However, our analysis reveals that this conclusion depends sensitively on the symmetry tolerance used in the structural characterization. Using strict numerical tolerances, structural relaxation introduces small distortions ($\sim 10^{-3}$~\AA) that lower the detected symmetry. In contrast, when more relaxed tolerances are employed ($\text{pos-tol} \sim 10^{-1}$), the system recovers the same $P2_1/c$ symmetry as in the bulk, with Cu atoms occupying equivalent Wyckoff positions. This result indicates that the monolayer remains very close to the bulk structural prototype, and that the apparent symmetry reduction originates from minor relaxation-induced distortions rather than from a fundamental symmetry breaking. 

 CuF$_2$ has emerged as a prototypical altermagnetic material, where non-relativistic 
spin splitting originates from the interplay of magnetic multipoles and lattice 
distortions~\cite{bandyopadhyay2026d}. Reducing the dimensionality introduces 
additional degrees of freedom: ferroelastic switching and stacking order in bilayers 
enable active control over the altermagnetic spin texture~\cite{peng2025ferroelastic,
peng2026sliding}. Similar to the already reported in other AM bilayer candidates such as MnPS$_3$ ~\cite{sun2025proposing,gonzalez2025engineering}.

\subsection*{Altermagnetism,\! Altermagnons and Cycloidal Phonons}

Figure~\ref{fig:3}(a) shows the non-relativistic electronic structure of CuF$_2$. 
The altermagnetic band splitting exhibits a strong dependence on the path in the Brillouin zone. 
Along the M$'$--$\Gamma$--M and X$'$--Y--X directions, a clear momentum-dependent spin splitting is observed, in agreement with previous studies that classify CuF$_2$ as a $d$-wave altermagnet. In contrast, along the X--T--X path, the bands preserve spin degeneracy.\\

\begin{figure}[ht] 
{\includegraphics[width=0.5\textwidth]{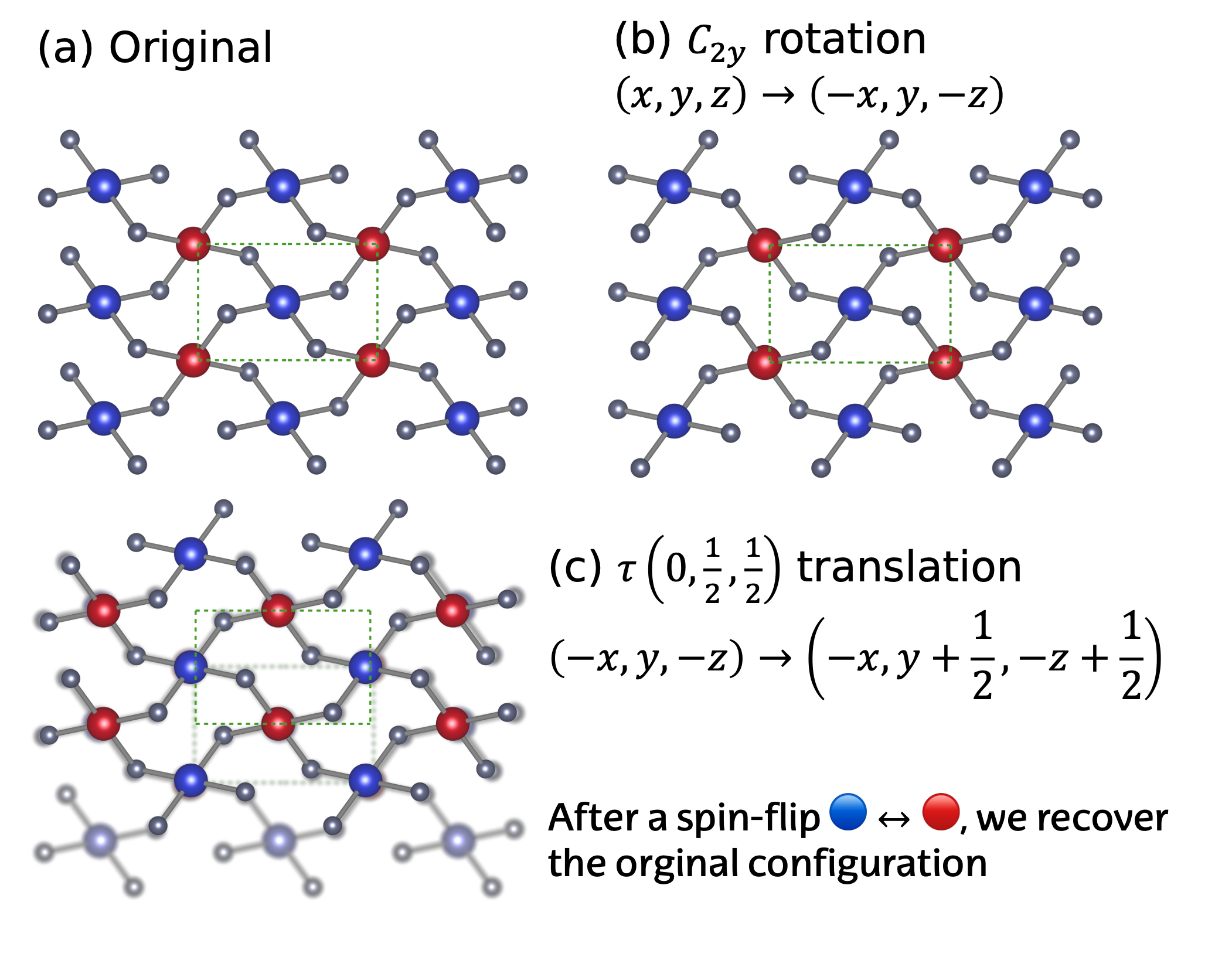}}
    \caption{Symmetry operations leading to the altermagnetic character: (a) Original structure. (b) and (c) Rotation and translation operations required to connect the spin-up and spin-down sublattices.}
    \label{fig:4}
\end{figure}

The altermagnetic order in the monolayer 
CuF$_2$ is governed by four symmetry operations of the spin space group 
$P2_1/c$, which in Seitz notation reads
$\{1 \mid 1 \mid \boldsymbol{\tau}(0,0,0)\}$,
$\{\bar{1} \mid m_{010} \mid \boldsymbol{\tau}(0,\frac{1}{2},\frac{1}{2})\}$,
$\{\bar{1} \mid 2_{010} \mid \boldsymbol{\tau}(0,\frac{1}{2},\frac{1}{2})\}$,
and $\{1 \mid \bar{1} \mid \boldsymbol{\tau}(0,0,0)\}$,
where the first entry denotes the spin rotation and the second the spatial 
operation. The key operation that exchanges the two spin-opposite Cu 
sublattices is $\{\bar{1} \mid 2_{010} \mid \boldsymbol{\tau}\}$, which 
combines a spin reversal with a twofold rotation $C_{2y}$ about the $b$ axis 
and a fractional translation, acting in real space as 
$(x,y,z) \rightarrow \left(-x, y+\tfrac{1}{2}, -z+\tfrac{1}{2}\right)$ (see Fig.~\ref{fig:4}).

In reciprocal space, the anisotropic action of these four 
operations is direction-selective: along $\Gamma$--M, where no operation 
maps $\mathbf{k}$ onto itself while simultaneously exchanging spin sectors, 
momentum-dependent spin splitting is symmetry-allowed; along $\Gamma$--X,  by contrast, the non-symmorphic operations enforce spin degeneracy by 
mapping $\mathbf{k}$ onto itself modulo a reciprocal lattice vector. This  directional selectivity is the symmetry origin of the altermagnetic band 
structure and, as we show below, simultaneously dictates the phonon angular 
momentum.

The transformation properties of collective excitations (phonons and magnons) are further constrained by the mirror symmetry $M_y$. For phonons, this operation reverses the sign of the in-plane phonon angular momentum ($J_z \rightarrow -J_z$), while for magnons it contributes to the directional modulation of the Dzyaloshinskii--Moriya-induced magnon chirality along different Brillouin zone directions. Together, these symmetry operations establish the common symmetry framework underlying the electronic, magnonic, and phononic responses of monolayer CuF$_2$.


In the case of phonons, the angular momentum does not inherit the momentum-space structure of the altermagnetic splitting. Rather, both phenomena originate from the same symmetry-imposed constraints, which dictate their distinct directional behavior, as we discuss in the next section. In two dimensions, the phonon angular momentum is given by
\begin{equation}
J_z \sim u_x \dot{u}_y - u_y \dot{u}_x,
\end{equation}
which characterizes the in-plane rotational motion of the lattice.
As shown in Fig.~\ref{fig:3}(b), a finite phonon angular momentum signal emerges along the X--$\Gamma$--X$'$ direction, which coincides with the path with spin-degeneracy identified in the electronic band structure (see Fig.~\ref{fig:3}(a)). The X--$\Gamma$--X$'$ path exhibits spin degeneracy in the absence of SOC. However, these $k$-points are not time-reversal-invariant points, and time-reversal symmetry is still broken; therefore, the phonon angular momentum can be observed in this region of the k-space.
To characterize its longitudinal component, we evaluate the helicity, given by the projection $\mathbf{J}\cdot\hat{\mathbf{k}}$. In this case, $\mathbf{J}\cdot\hat{\mathbf{k}} = 0$, indicating that the phonon angular momentum has no longitudinal component. The phonon modes exhibit cycloidal motion \cite{juraschek2025chiral}. It is worth mentioning that the Néel ground state breaks both time-reversal symmetry and the inversion center of the monolayer, which constitutes the key ingredient for activating the phonon angular momentum. By performing the same calculations for the FM configuration, we find that $J_z = 0$, indicating that the phonon angular momentum is suppressed in the centrosymmetric ferromagnetic phase.
We now turn to the magnon features. When SOC is included, the full spin space group
symmetry $P2_1/c\,1$ 
is reduced to the magnetic
space group $P\bar{1}$, Type~I collinear altermagnet \cite{cheong2024altermagnetism}.
At generic $k$-points, the MSG little co-group contains only the
identity, enabling momentum-dependent spin splitting throughout the BZ.
The magnetic configuration remains collinear, with antiferromagnetic
order lying in the plane defined by the in-plane lattice vectors.
The Néel vector is oriented approximately along the $[\bar{1}10]$
crystallographic direction (equivalent to $[-\sqrt{2}/2, \sqrt{2}/2, 0]$ in the Cartesian spin frame), consistent with the collinear direction
reported by the \textit{FindSpinGroup} analysis in both the Cartesian
and oriented spin frames.

In the low-symmetry $P\bar{1}$ setting, the two Cu sites occupy
distinct Wyckoff positions (1$f$ and 1$g$) in the magnetic space
group, which are symmetry-inequivalent and not mapped onto each other by any operation of the MSG. This allows anisotropic exchange
interactions, including Dzyaloshinskii-Moriya terms. 
Our first-principles calculations yield an isotropic exchange
$J_1 \approx -12.35~\mathrm{meV}$ that sets the dominant
antiferromagnetic energy scale, while the symmetric anisotropic
exchange ($|J_{zz}| \approx 0.87~\mathrm{meV}$, $\sim 7\%$ of $|J_1|$)
selects the spin-ordered plane. The Dzyaloshinskii--Moriya interaction
($|\mathbf{D}_1| \approx 0.33~\mathrm{meV}$, $\sim 3\%$ of $|J_1|$) governs the weak ferromagnetic canting observed along $[\bar{1}\bar{1}0]$, perpendicular to the Néel vector within the plane. 

Fig.~\ref{fig:3}(c) shows the magnon spectrum, which exhibits two eigenmodes 
associated with the two magnetic sublattices and well-defined minima at 
high-symmetry points. As shown in Fig.~\ref{fig:3}(c), the chiral 
response is strongly anisotropic, being concentrated along the 
$M^\prime -\Gamma - M$ direction rather than distributed throughout 
the Brillouin zone. This corresponds to the antinodal altermagnetic path and 
is consistent with previous studies showing that the same symmetry operations responsible for the altermagnetic spin splitting also govern the chirality-split 
magnon spectra~\cite{vsmejkal2023chiral}.

To elucidate the microscopic origin of this anisotropic response, we selectively switched off the different interaction terms in Eq.~(1), as 
shown in Fig.~S1 of the Supplemental Material. This term-resolved analysis 
reveals that the symmetric anisotropic exchange $J_{\mathrm{ani}}$, which 
provides the microscopic counterpart of the phenomenological Anisotropic 
Altermagnetic Stiffness (AAS) in the continuum limit~\cite{gomonay2024structure}, 
drives the dominant reconstruction of the magnon spectrum, producing a 
maximum energy shift of approximately $208$~meV along the 
$M^\prime-\Gamma-M$ direction. In contrast, the 
Dzyaloshinskii--Moriya interaction contributes only a secondary correction 
of about $0.5$~meV, mainly localized near the $\Gamma$ point in the lower 
magnon branch. The large ratio between the corresponding maximum spectral 
shifts,
$\Delta E_{J_{\mathrm{ani}}}^{\max}/\Delta E_{\mathrm{DMI}}^{\max}
\approx 40$, demonstrates that the altermagnetic magnon response in 
CuF$_2$ is primarily governed by symmetric anisotropic exchange, while the 
DMI acts as a weak, direction-dependent modulation superimposed on this 
altermagnetic exchange baseline.

Along the $X^\prime-Y-X$ path, by contrast, the 
high-energy magnon branches remain nearly degenerate regardless of whether 
DMI or $J_{\mathrm{ani}}$ is retained. This behavior indicates that the suppression is not a fine-tuned consequence of a particular interaction, but 
rather reflects a symmetry-constrained reduction of the anisotropic exchange 
projected onto these modes. In particular, the effective projection of 
$J_{\mathrm{ani}}$ onto the observable high-energy magnon splitting is 
strongly suppressed along this direction, even though $J_{\mathrm{ani}}$ 
remains the dominant anisotropic term in the global spectrum. This 
directional selectivity, maximal response along 
$M^\prime-\Gamma-M$ and suppressed response along $X^\prime-Y-X$, is consistent with the phenomenology of 
$d$-wave altermagnets, where the AAS contribution is maximal along 
antinodal directions and suppressed along nodal directions by 
symmetry~\cite{gomonay2024structure}.

In contrast to the nodal $X^\prime-Y-X$ direction, the
$M^\prime-\Gamma-M$ path displays the maximal magnon response.
Along this path, the spectrum remains gapped except for a narrow reduced-gap
region near $\Gamma$. This region coincides with a large chiral component
$\chi_{AB}(\mathbf{k})$, as shown in Fig.~\ref{fig:3}(c). It therefore
marks the momentum-space region where the relative phase structure of the
magnon Bogoliubov amplitudes evolves most strongly. Importantly, this
redistribution is not merely a spectral feature, but a manifestation of the
geometric and symmetry-imposed structure of the magnon bands.

The same altermagnetic symmetry that governs the momentum-dependent spin splitting also constrains the phase structure of the magnon wavefunctions. This nontrivial phase evolution generates a finite Berry curvature that is highly localized and anisotropic, with dominant contributions concentrated near the reduced-gap avoided crossings. Its momentum-space distribution exhibits a dipolar-like pattern that faithfully mirrors the underlying $d$-wave altermagnetic symmetry.

Numerical integration over the discretized Brillouin zone yields quantized magnon Chern numbers $C^M=\pm 2$, confirming the non-trivial topology of the magnon bands~\cite{khatua2025magnon}. This quantization corresponds to a cumulative Berry-phase winding of $4\pi$ across the Brillouin zone and is enabled by the symmetry reduction to the magnetic space group $P\bar{1}$ upon including spin--orbit coupling. In this low-symmetry magnetic setting, the anisotropic exchange-driven phase texture is allowed to generate a net Berry flux. Consequently, the nonzero magnon Chern numbers establish monolayer CuF$_2$ as a two-dimensional altermagnetic platform for topological magnon transport. In contrast to conventional topological magnon insulators typically relying on ferromagnetic order or noncollinear spin textures, this phase emerges from a collinear, fully compensated magnetic state, identifying altermagnetism as a symmetry-protected route to chiral bosonic excitations. The resulting bands are expected to host chiral magnon edge modes and a finite transverse thermal Hall conductivity, providing direct experimental signatures of the underlying altermagnetic topology~\cite{khatua2025magnon,AgF2_paper}.


Now we focus on the phonon angular momentum, which exhibits a behavior distinct from that of magnons. While the magnonic response is maximized along the altermagnetic axis, the phonon angular momentum instead emerges along the $\Gamma$–$X$ direction. This distinction originates from the transformation properties of the atomic displacements under the same symmetry operations, given by $\{E, P, t_{1/2}M_y, t_{1/2},C_{2y}\}$ of the $P2_1/c$ space group.\\

Under the mirror operation $M_y$, the in-plane displacement components transform as $(u_x, u_y) \rightarrow (-u_x, u_y)$, and similarly for the velocities, yielding $J_z \rightarrow -J_z$. An analogous transformation follows for the twofold rotation $C_{2y}$, whose restriction to the $(x,y)$ plane also reverses the orientation (determinant $-1$), leading again to $J_z \rightarrow -J_z$. More generally, $J_z$ changes sign under any symmetry operation that reverses the orientation of the plane.
As a consequence, a finite phonon angular momentum can only arise along directions where no symmetry operation leaves $\mathbf{q}$ invariant while simultaneously reversing $J_z$. This symmetry constraint suppresses the phonon chirality along the altermagnetic $\Gamma$–$M$ path, while allowing for a finite cycloidal phonon response along $\Gamma$–$X$, thereby revealing a complementary relationship between altermagnetic electronic structure and lattice angular momentum.


In particular, along the $\Gamma$--X path, neither $C_2$ nor $M_y$ imposes $J_z = 0$: $C_2$ enforces a spin-degenerate nodal line but leaves $J_z$ unconstrained, while $M_y$ maps $(q_x,0)\to(-q_x,0)$, yielding $J_z(q_x)=-J_z(-q_x)$, which relates opposite momenta without imposing a local constraint at any given $\mathbf{q}$. A finite cycloidal phonon angular momentum is thus symmetry-allowed along this direction.

In contrast, along $\Gamma$--M, no nontrivial point symmetry leaves the wave vector invariant, reducing the little group to the identity. Symmetry operations such as $M_y$ map $(q,q)\to(-q,q)$ and relate $J_z(q,q)$ to $-J_z(-q,q)$, again connecting symmetry-related momenta without enforcing a local constraint. Accordingly, while the absence of symmetry restrictions permits a finite nonrelativistic spin splitting, the phonon angular momentum is nevertheless found to vanish. This dual behavior reflects the global $d$-wave texture of the system: the $\Gamma$--X path probes regions where $[C_{2} \parallel C_{2}^{\mathrm{spin}}]$ stabilizes phonon chirality and symmetry-protected electronic nodal lines, whereas $\Gamma$--M traverses nodal regions of the alteraxial texture, where spin splitting is allowed and the phonon angular momentum vanishes. \\


This complementary directional selectivity contrasts with recent 
reports on helically ordered altermagnets, where magnetic helicity fosters the coexistence and mutual enhancement of chiral magnons and chiral phonons 
along the screw axis~\cite{okamoto2025altermagnetic}. In CuF$_2$, however, the same 
non-symmorphic symmetries that govern the altermagnetic magnon response 
produce a complementary momentum-space separation of spin and lattice 
chirality: cycloidal phonon angular momentum emerges precisely along 
directions where the magnon chiral response is symmetry-suppressed. This 
highlights a distinct, symmetry-imposed regime of spin-lattice chirality 
in a collinear two-dimensional altermagnet.

\section*{Final Remarks}

We have demonstrated that monolayer CuF$_2$ realizes a unified symmetry origin for spin and lattice chirality, where the same nonsymmorphic operations governing altermagnetic spin splitting simultaneously control chiral magnons and cycloidal phonon angular momentum while directing them into complementary regions of momentum space. The magnon spectrum is predominantly shaped by symmetric anisotropic exchange, the microscopic counterpart of the Anisotropic Altermagnetic Stiffness, whereas the Dzyaloshinskii--Moriya interaction plays only a secondary role. This identifies collinear nonsymmorphic symmetry as a distinct SOC-free route to coupled spin--lattice chirality, fundamentally different from the DMI-driven mechanism of helical altermagnets~\cite{okamoto2025altermagnetic}. Remarkably, spin--orbit coupling does not generate the chirality itself, but lowers the symmetry to $P\bar{1}$ and gives rise to quantized Chern numbers $C^M=\pm2$, establishing an altermagnetic topological magnon phase in a fully compensated collinear magnet. Chiral magnon excitations have recently been reported in closely related fluoride altermagnets such as FeF$_2$ \cite{sears2026altermagnetic} and MnF$_{2}$, highlighting the growing experimental interest in chirality-resolved spin dynamics \cite{faure2025altermagnetism}.

Moreover, the $P2_1/c$ spin space group corresponds
to the $2/m$ crystallographic point group, which in
the classification of Zhang \textit{et al.}~\cite{zhang2026oddparity}
is compatible with odd-parity magnon splitting once the effective time-reversal symmetry is broken.
In conventional antiferromagnets, this symmetry
must be broken externally, for example, by circularly
polarized light; in monolayer CuF$_2$, however, the
altermagnetic order already breaks it intrinsically,
so that external driving could superimpose a tunable
odd-parity component directly onto the pre-existing
altermagnetic baseline, potentially enabling
Floquet-controlled transitions between topologically
distinct magnon phases. The coexistence of intrinsic
cycloidal phonons with finite angular momentum along
$\Gamma$--X further suggests that chiral-phonon-assisted
Floquet schemes~\cite{zhang2026oddparity} could offer a
symmetry-selective pathway to modulate magnon
chirality via indirect spin--lattice coupling,
bypassing the weak bare Aharonov--Casher interaction.
Monolayer CuF$_2$ thus connects non-relativistic
altermagnetic order and relativistic topological
responses through a single non-symmorphic symmetry framework that simultaneously governs both.

These findings open several concrete directions. In bilayer 
geometries, layer sliding and twisting could provide additional 
degrees of freedom to tune the altermagnetic domain and thereby 
select or suppress specific chiral phonon branches, suggesting 
a phononic analogue of spin-layertronics. More broadly, the directional complementarity between magnon and phonon chirality identified in CuF$_2$ suggests a symmetry-based route for designing coupled spin--lattice chiral responses. In this system, momentum directions that host symmetry-enforced nodal lines in the magnon spectrum coincide with directions where phonon angular momentum is symmetry-allowed. These results motivate the search for similar phenomena in other two-dimensional altermagnetic systems and related magnetic materials.

\section*{Acknowledgments}
C. A. was supported by the Foundation for Polish Science project “MagTop” no. FENG.02.01-IP.05-0028/23 co-financed by the European Union from the funds of Priority 2 of the European Funds for a Smart Economy Program 2021–2027 (FENG). Powered@NLHPC: This research was partially supported by the supercomputing infrastructure of the NLHPC (CCSS210001). A.L. acknowledges support from ANID through FONDECYT Iniciación Grant No. 11251906.

\bibliographystyle{unsrtnat}
\bibliography{bib}
\newpage
\pagebreak

\onecolumn
\begin{center}

\section*{Supplementary Information}

\textbf{\large{Chiral Magnons and Cycloidal Phonons in Altermagnetic CuF$_{2}$ monolayers}}

\end{center}

\setcounter{figure}{0} 
\setcounter{section}{0} 
\setcounter{equation}{0}
\setcounter{table}{0}
\setcounter{page}{1}
\renewcommand{\thepage}{S\arabic{page}} 
\renewcommand{\thesection}{S\Roman{section}}   
\renewcommand{\thetable}{S\arabic{table}}  
\renewcommand{\thefigure}{S\arabic{figure}} 
\renewcommand{\theequation}{S\arabic{equation}} 

To clarify the role of the different magnetic interactions entering the spin Hamiltonian of Eq.~(1), we performed a series of calculations in which selected terms were systematically removed. Figure~S1 compares the resulting magnon spectra with the full model. Panel (a) shows the magnon band structure obtained by including all interactions. In panel (b), the Dzyaloshinskii--Moriya interaction is set to zero ($D=0$), allowing us to assess its contribution to the magnon dispersion. Panel (c) shows the effect of suppressing the anisotropic exchange interaction ($J_{\mathrm{ani}}=0$), while panel (d) presents the spectrum obtained when both the Dzyaloshinskii--Moriya and anisotropic exchange terms are neglected.

\begin{figure*}[ht]
\centering

\subcaptionbox{}{\includegraphics[width=0.4\textwidth]{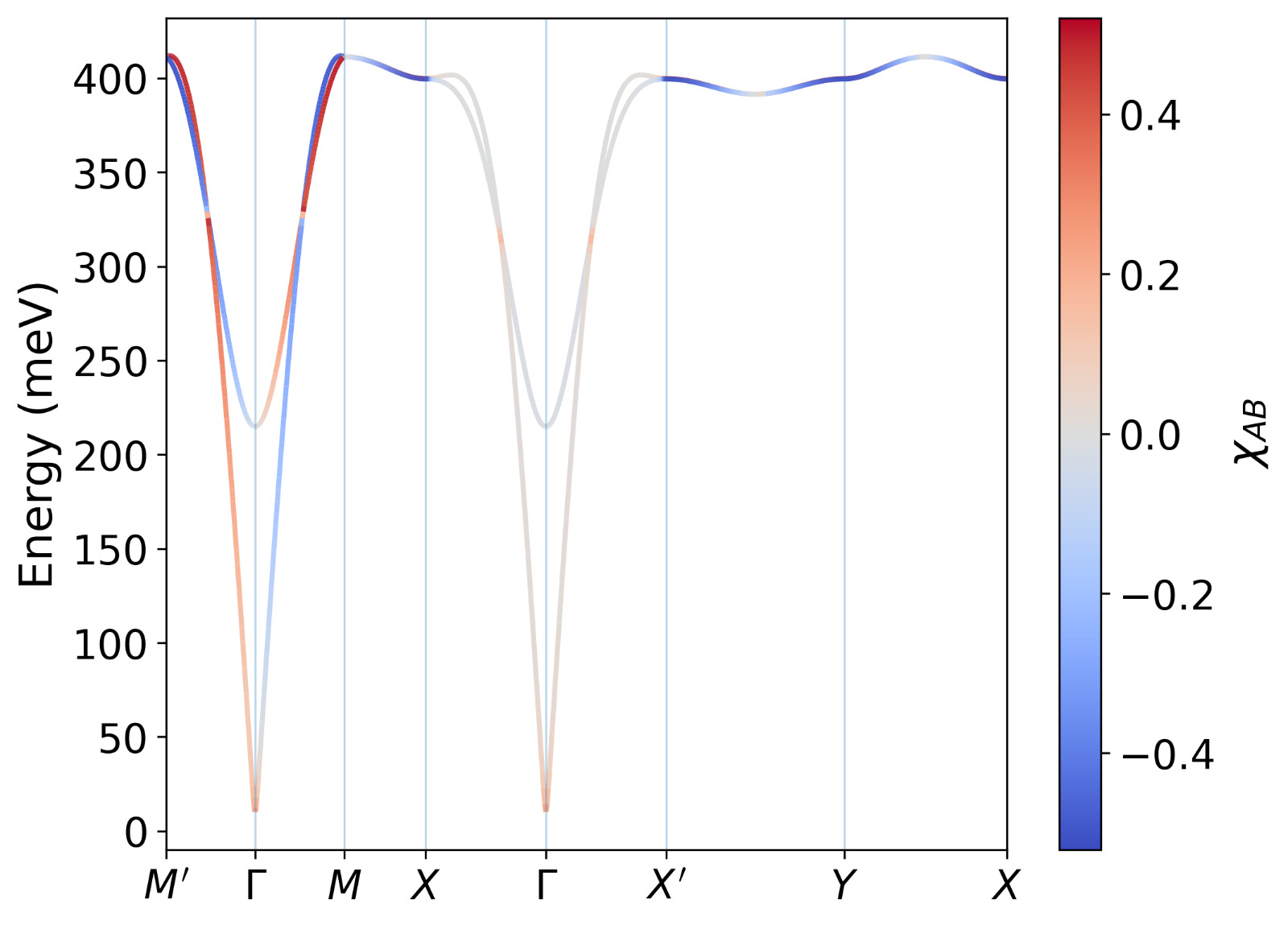}}
\subcaptionbox{}{\includegraphics[width=0.4\textwidth]{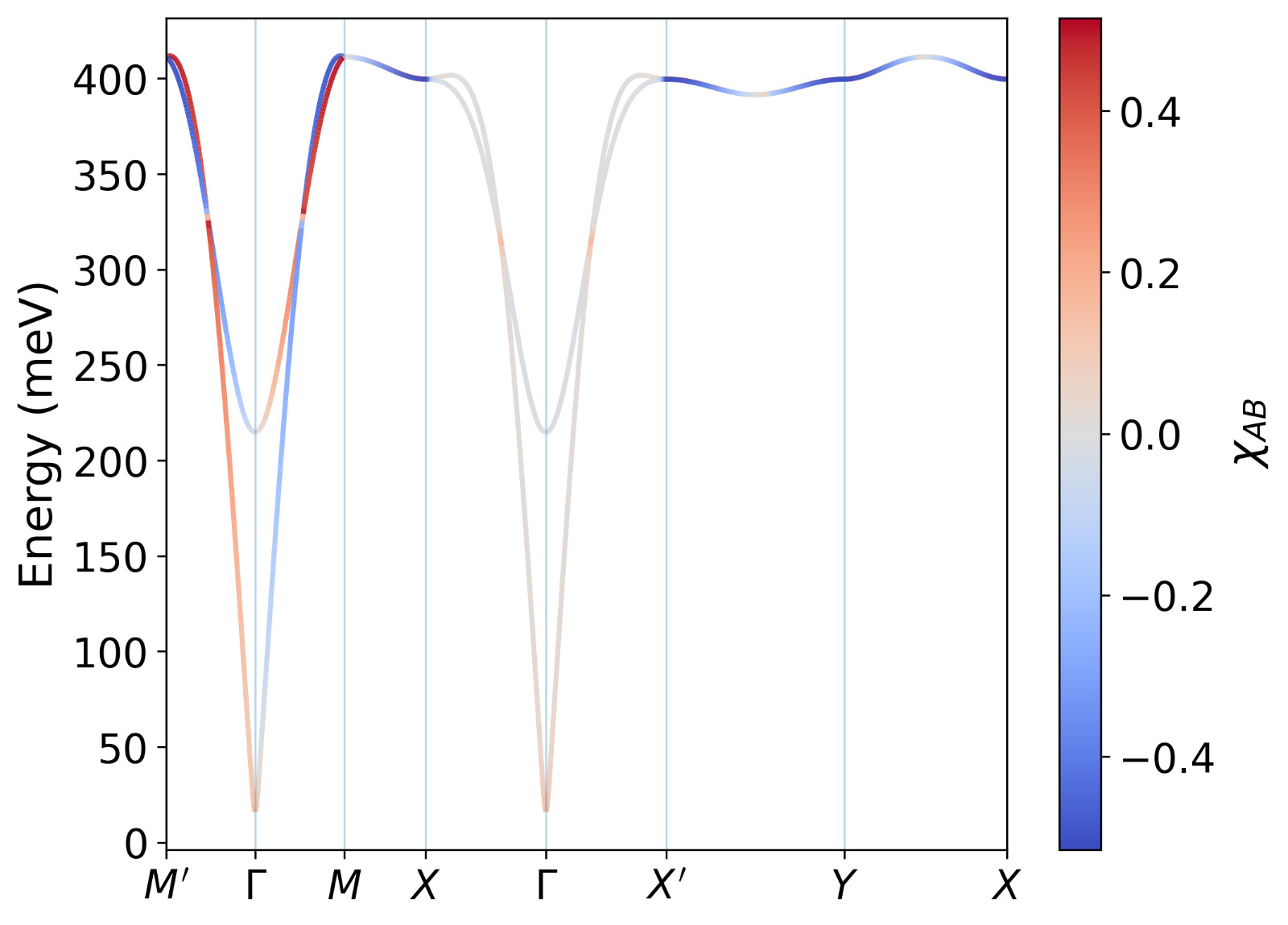}}

\subcaptionbox{}{\includegraphics[width=0.4\textwidth]{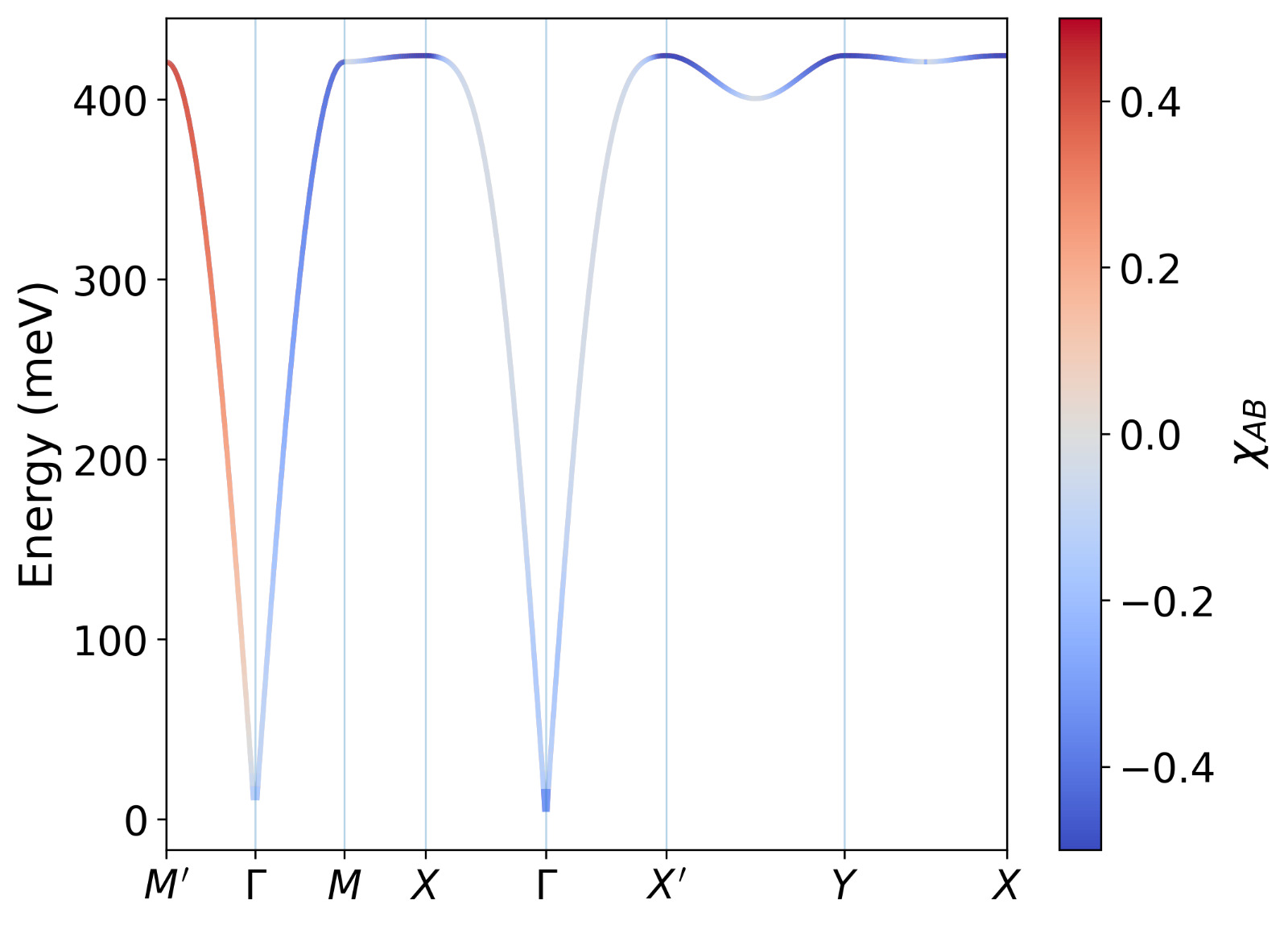}}
\subcaptionbox{}{\includegraphics[width=0.4\textwidth]{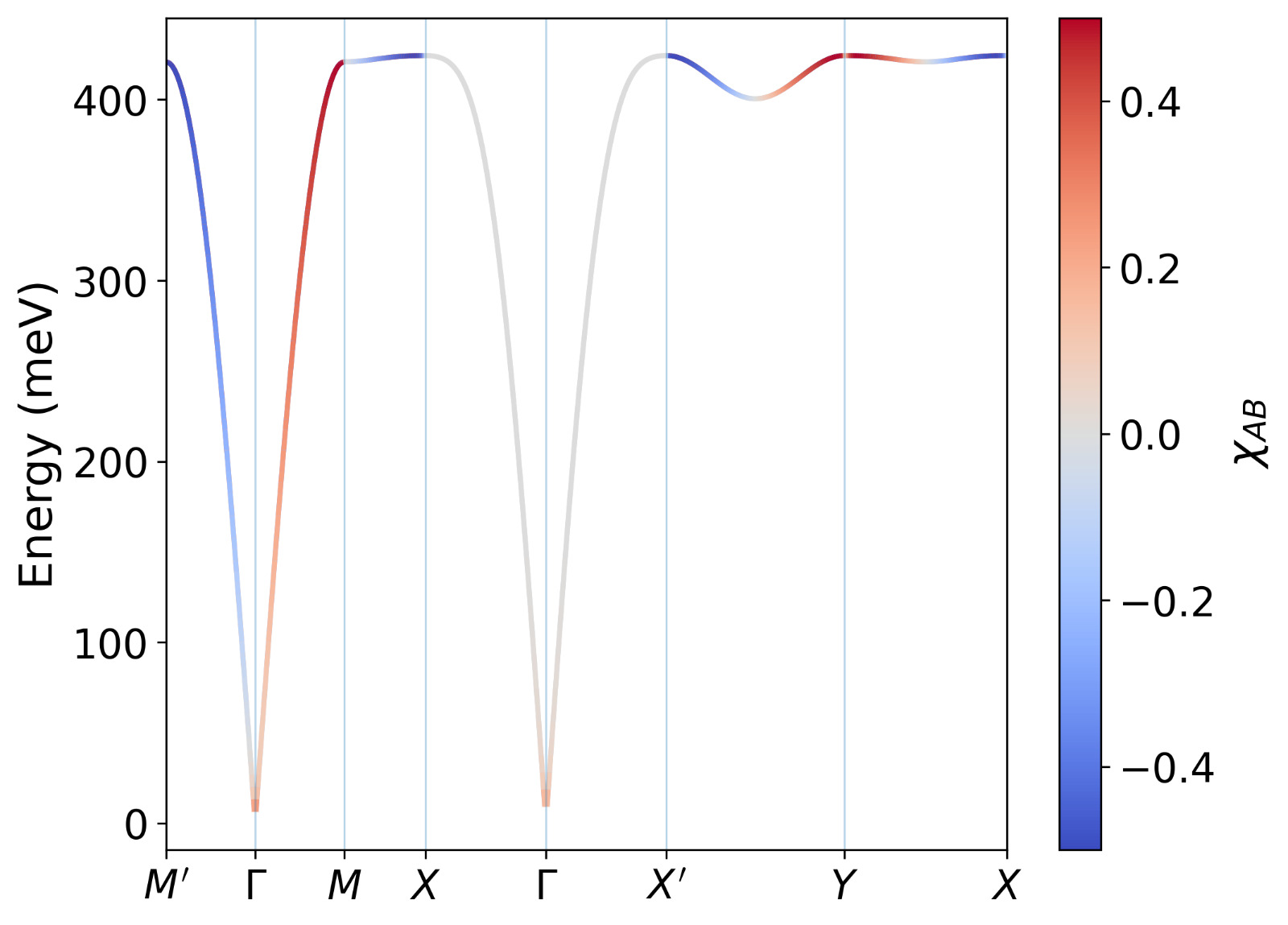}}

\caption{
Magnon band structure: (a) Considering all Hamiltonian terms.
(b) Switching off the Dzyaloshinskii--Moriya interaction terms ($D=0$).
(c) Switching off the anisotropic exchange terms ($J_{\mathrm{ani}}=0$).
(d) Switching off both the Dzyaloshinskii--Moriya and anisotropic exchange terms ($D=J_{\mathrm{ani}}=0$).
The color scale represents the magnitude and sign of the phonon angular momentum $J_z$ and the inter-sublattice coherence $\chi_{AB}$, inter-sublattice coherence $\chi_{AB}$, which quantifies the degree of hybridization between the $A$ and $B$ sublattices. Red (blue) indicates negative (positive) values of $\chi_{AB}$.}
\label{fig:S1}
\end{figure*}

\textbf{\large{}}

\setcounter{figure}{0} 
\setcounter{section}{0} 
\setcounter{equation}{0}
\setcounter{table}{0}
\setcounter{page}{1}
\renewcommand{\thepage}{S\arabic{page}} 
\renewcommand{\thesection}{S\Roman{section}}   
\renewcommand{\thetable}{S\arabic{table}}  
\renewcommand{\thefigure}{S\arabic{figure}} 
\renewcommand{\theequation}{S\arabic{equation}} 

\end{document}